\documentclass[12pt,preprint]{aastex}

\usepackage{graphicx}

\newcommand{\beq}{\begin{equation}}
\newcommand{\eeq}{\end{equation}}
\newcommand{\beqa}{\begin{eqnarray}}
\newcommand{\eeqa}{\end{eqnarray}}
\newcommand{\pp}{^{\prime\prime}}

\def\la{\lower.5ex\hbox{$\; \buildrel < \over \sim \;$}}
\def\ga{\lower.5ex\hbox{$\; \buildrel > \over \sim \;$}}

\begin{document}

\title{Orbit of the Large Magellanic Cloud in a \\ Dynamical Model for the Local Group}

\author{P.~J.~E. Peebles}  
\affil{Joseph Henry Laboratories, Princeton University, Princeton, NJ 08544, USA}

\begin{abstract}
A mass model that includes galaxies in and near the Local Group and an external mass in the direction of the Maffei system, with the condition from cosmology that protogalaxies have small peculiar velocities at high redshifts, allows a plausible picture for the past motion of the Large Magellanic Cloud relative to the Milky Way. The model also fits the proper motions of M33  and IC10.
\end{abstract}
\maketitle

\section{Introduction}\label{sec:1}

Analyzing the past orbit of the Magellanic Clouds relative to the Milky Way is an interesting problem in dynamics, and it may be a useful preliminary exercise for analyses of ongoing advances in measurements of galaxy distances and proper motions that will test ideas about the evolving mass distribution around galaxies. Since the positions, velocities and accelerations of the Magellanic Clouds are fairly well constrained at low redshifts (Kallivayalil {\it et al.} 2006, Piatek, Pryor \& Olszewsk 2008), the challenge here is to understand how the Clouds were directed to their present paths by interactions among newly forming galaxies at high redshift. This analysis of the motion of the Large Magellanic Cloud (LMC) around the Milky Way (MW) focuses on the influence of galaxies in and near the Local Group. The Small Cloud is taken to be an unimportant perturbation at the hoped for accuracy of this study.  

The first steps to the interpretation of the measured proper motion of the Large Magellanic Cloud considered its interaction with MW and M31 (Besla  et al. 2007; Shattow  \& Loeb 2009; Kallivayalil {\it et al.} 2009). Peebles (2009, P9) added the initial condition indicated by the gravitational growth of structure in the standard Big Bang cosmology, that peculiar velocities of the protogalaxies at high redshift are growing. P9 applied this condition to a illustrative mass model. Here the analysis is extended to a more realistic model that takes account of the accepted picture of the mass structure in the Local Group (LG). Sensitivity to the mass outside the LG  is explored by comparing solutions with and without a large external mass near IC342 and Maffei 1. 

The numerical method grew out of the action approach in Peebles (1989) for dealing with the mixed boundary conditions that initial peculiar motions are small and present positions of the galaxies are consistent with what is observed. This paper introduces a method of solution that improves the efficiency of the numerical action method enough to suggest a new name, NNAM. The method of solution is described in Section \ref{sec:2} and the Appendix.

Since the computation is approximate --- the broadly distributed mass in a galaxy is represented by a single particle, and the model deals with only a few galaxies --- one cannot expect to achieve an exact fit to the galaxy positions and motions. To deal with this, the approach taken here is to seek orbits that minimize a $\chi^2$ measure of fit to the physical parameters with assigned uncertainties that are treated as standard deviations. The physical parameters include the proper motions of LMC, M33 and IC10, where the stated measurement  uncertainties are treated as standard deviations. The circular velocity of MW is assigned what seems to be a reasonable measurement uncertainty. The galaxy masses are assigned central values and uncertainties, as in  Shaya \& Peel (2007), that are meant to represent the ranges of what are generally considered to be reasonable values. The present positions and redshifts of the galaxies are allowed  uncertainties larger than the measurement errors. This may may be taken to represent an actual offset of the center of starlight from the effective center of the mass that is dominated by the dark matter on the outskirts. For example, if the dark matter 100~kpc from the center of M31 were as irregularly distributed as the stars (McConnachie et al. 2009) then it would be easy to imagine a substantial offset of the effective mass center of this galaxy from the center of its stars. But these offsets in galaxy positions and motions may just as well be considered a way to allow for the approximate nature of the model. The parameters and their assigned uncertainties for  this procedure are discussed in Section  \ref{sec:3}.

This approach allows considerable freedom of adjustment of parameters, but there is a considerable number of constraints. The results in Section~\ref{sec:4} include discovery of just one form of orbits, within modest variations, that offers a reasonable fit to all the parameters, including the proper motions of LMC, M33 (Brunthaler et al. 2005) and IC10 (Brunthaler et al. 2007). I argue that the results offer a good case for the shape of the orbit of LMC relative to MW back to high redshift. The motion of M31 requires more work, and since M33 and IC10 are affected by what M31 has been doing the orbits presented here for M33 and IC10 are best considered examples of what might have happened, to be checked by more ambitious models and still better galaxy distances and proper motions. 
 
\section{Numerical Methods}\label{sec:2}

This section reviews the dynamical model and the  numerical action procedure. The details of the new method of solution are described in the Appendix. 

 The starting idea is to reduce the evolution of the mass distribution to an N-body problem in which a single particle represents the mass concentrated around a galaxy or a tightly bound system of galaxies. This is thought to be a good approximation at low redshift because the galaxies in our neighborhood by and large are well separated relative to standard  estimates of the sizes of their dark matter halos. The approximation need not be vitiated by merging, because a particle may trace the effective mass center and momentum of the two systems before the merger as well as in the merged system. Further details of this line of argument may be traced back through P09. 

In this analysis of the N-body problem the Friedmann-Lema\^\i tre cosmology is taken to be spatially flat with constant dark energy density. The expansion parameter $a(t)$ satisfies 
\beq
{\dot a^2\over a^2} = {H_o^2\Omega\over a^3} + (1 - \Omega)H_o^2,
\label{eq:FLeq}
\eeq
with present value $a_o=1$. Hubble's constant is $H_o$, $\Omega$ is the density parameter, and matter pressure is ignored.

The action 
\beq
S=\int_0^{t_{o}}dt\left[\sum_i {m_i a^2 \dot x_i^2\over 2} +
{1\over a}\left(-\sum_{j\not= i}\phi_{i,j} +
{1\over 4}\sum_i m_i\Omega H_o^2 x_i^2\right)\right],
\label{eq:action}
\eeq
summed over particles $i$, with fixed present positions and the initial condition
\beq
a^2{d\vec x_i\over dt}\rightarrow 0 \hbox{ at } a(t)\rightarrow 0,
\label{eq:initialcondition}
\eeq
gives the equation of motion
\beq
{d\over dt}a^2{d\,x_{i,k}\over dt}= a\left({\tt g }_{i,k} +{\Omega H_o^2x_{i,k}\over 2a^2}\right).
\label{eq:eqofm}
\eeq
The Cartesian coordinate label is $k=1, 2, 3$.  Physical lengths are $r_{i,k} = ax_{i,k}$, and ${\tt g}_{i,k}=g_{i,k}/a^2$ is the physical acceleration of particle $i$ caused by the gravitational attraction of the other particles. Since the cosmic mean mass density  is $\bar\rho(t)=3\Omega H_o^2/(8\pi Ga(t)^3)$, the second term in parentheses on the right hand side of equation~(\ref{eq:eqofm}) is the physical acceleration $4\pi G\bar\rho\, r_{i,k}/3$. This counter term eliminates the accelerations of the particle coordinate positions $x_{i,k}$ when the coordinate positions are at rest and arranged so their mutual gravitational accelerations ${\tt g}_{i,k}$ are the same as that of the background cosmology.

In a discrete time step approximation the equation of motion (\ref{eq:eqofm}) is 
\beqa
0=S_{i,k,n} &=& -{a_{n+1/2}^2\dot a_{n+1/2}\over a_{n+1} - a_n}(x_{i,k,n+1}-x_{i,k,n}) \nonumber\\
&+& {t_{n+1/2}-t_{n-1/2}\over a_n}
\left[\sum_{j\not=i}g_{i,j,k,n} +
{1\over 2}\Omega H_o^2 x_{i,k,n}\right]\label{eq:discrete_eom}\\
&+& {a_{n-1/2}^2\dot a_{n-1/2}\over a_{n} - a_{n-1}}(x_{i,k,n}-x_{i,k,n-1}). \nonumber
\eeqa
The coordinate acceleration is here written as the sum over the pull of all other particles, and it will be recalled that the physical acceleration is ${\tt g}_{i,j,k} =g_{i,j,k}/a^2$. The indices are Cartesian coordinates $k = 1,2,3$, $1\leq i \leq$ the particle number, and time steps $1\leq n \leq n_x+1$. The expansion parameter $a_{n+1/2}$ and time  $t_{n+1/2}$ are evaluated half-way between time steps $n$ and $n+1$, producing a leapfrog approximation to the equation of motion. In the action method the present positions $x_{i,k,n_x+1}$ are given and the initial condition in equation~(\ref{eq:initialcondition}) is represented by setting $a_{1/2}=0$ so that equation~(\ref{eq:discrete_eom}) at $n=1$ is
 \beq
0 = S_{i,k,1} = -{a_{3/2}^2\dot a_{3/2}\over a_{2} - a_1}(x_{i,k,2}-x_{i,k,1}) 
+ {t_{3/2}\over a_1}\left(\sum_{j\not=i}g_{i,j,k,1} +
{1\over 2}\Omega H_o^2 x_{i,k,1}\right).\label{eq:discrete_eom_i}
\eeq

When the free coordinates $x_{i,k,n}$ at $1\leq n\leq n_x$ are adjusted to make the same number of $S_{i,k,n}$ all vanish then the $x_{i,k,n}$ are a numerical solution in leapfrog approximation to the equation of motion (\ref{eq:eqofm}) for given final positions $x_{i,k,n_x+1}$ and the initial condition in equation~(\ref{eq:initialcondition}). One approach to $S_{i,k,n}=0$ is to walk down the the gradient of $S$ to a minimum (Peebles 1989), but that misses maxima and saddle points. The $S_{i,k,n}$ are driven to zero at these stationary points of the action by iteratively applying the coordinate shifts $\delta x_{i,k,n}$ from the solution to
\beq
S_{i,k,n} + \sum_{j,k',n'} {\partial S_{i,k,n}\over\partial x_{j,k',n'}}\delta x_{j,k',n'} = 0.
\label{eq:dreamon}
\eeq
Computing the $\delta x_{i,k,n}$ by matrix inversion, as in Peebles et al. (2001), is exceedingly slow for an interesting number of time steps. The NNAM described in the Appendix speeds this up by making use of the fact that in equation~(\ref{eq:discrete_eom}) the matrix 
$\partial S_{i,k,n}/\partial x_{j,k',n'}$ is nonzero only near the diagonal.

This analysis of the motion of LMC also uses numerical integration of the equation of motion (\ref{eq:eqofm}) forward in time from initial positions and proper peculiar velocities taken from a NNAM solution. These initial conditions are   
\beq
x_{i,k,3/2} = (x_{i,k,2}+x_{i,k,1})/2, \quad v_{i,k,3/2} = a_{3/2}\dot x_{i,k,3/2} =  
  a_{3/2}\dot a_{3/2}{x_{i,k,2}-x_{i,k,1}\over a_2 - a_1}. \label{eq:iv}
\eeq
This is accurate when the time steps are uniformly spaced in the expansion parameter $a(t)$ and the particle coordinates are varying in proportion to $a(t)$, as in the growing mode of departure from homogeneity of an ideal fluid in linear perturbation theory. It proves to be a good approximation in the numerical solutions presented here, as evidenced by the close agreement of present positions given to NNAM and present positions and velocities from forward numerical integration with these initial conditions. 

\section{Parameters}\label{sec:3}

\subsection{Numerical parameters}\label{sec:31}

To take account of the dark matter halos of MW and LMC when LMC approaches the MW at low redshift their mutual interactions are modeled on a rigid truncated limiting isothermal sphere, with coordinate acceleration
\beq
g_{\rm LMC} = {av_c^2\over x}, \quad
g_{\rm MW} = {m_{\rm LMC}\over m_{\rm MW}}{av_c^2\over x},\quad \hbox{for } ax < r_1 = {Gm_{\rm MW}\over v_c^2}, 
\label{eq:g_LMC}
\eeq 
where $m_{\rm MW}$ and $m_{\rm LMC}$ are the masses and $v_c$ is the LMC circular velocity at the Solar circle. This prescription conserves momentum. It ignores the perturbation to the MW mass distribution by LMC, as in dynamical drag, but P9 shows that the effect of dynamical drag on the motion of LMC is small in the LG model used here, because LMC has approached at high velocity. For other particles, and for the interaction of LMC and MW at larger separation, the coordinate acceleration of particle $i$ due to $j$ is the usual inverse square law form,  
\beq
g_{i,j,k} = Gm_j {x_{j,k}-x_{i,k}\over |x_i-x_j|^3}.
\label{eq:g}
\eeq
The MW mass structure is known in more detail than the model in equations~(\ref{eq:g_LMC}) and~(\ref{eq:g}), but this simplified scheme seems to be an appropriate match to the schematic nature of the mass model. 

The NNAM solutions have $n_x=500$ time steps, about as many as convenient computation time allows. The numerical integration forward in time from the initial conditions in equation~(\ref{eq:iv}) has $n_x=5000$ steps. In both cases the expansion parameter $a_n$ is uniformly spaced. The NNAM solutions start at $a_i=0.1$, or initial redshift 
\beq
1+z_i=10. \label{eq:zi} 
\eeq

In the iterative relaxation to a NNAM solution to the equation of motion the coordinates are shifted by $\epsilon\delta x_{i,k,n}$ where $\delta x_{i,k,n}$ is the solution to equation~(\ref{eq:dreamon}). At $\epsilon =1$ the iteration on occasion fails to converge. This is avoided by setting $\epsilon = 0.1$ far from a solution or where convergence is unusually slow. In the results presented here the orbits are relaxed to the equation of motion (\ref{eq:eqofm}) to near machine double precision.

The search for plausible NNAM solutions starts from random orbits (with parameters uniformly distributed over the ranges of nominal standard deviations in Table~1, orbits of LMC, M33 and IC10 that circle present positions $x_o$ as $x = x_o + A\sin(4 \pi B (1-a(t)))$, where $A$ is uniformly distributed between $\pm 5$~Mpc and $B$ is uniformly distributed between 0 and 1, and the other galaxies move linearly in $a(t)$ to present positions from  initial displacements uniformly distributed between $\pm 5$~Mpc in each Cartesian coordinate). If the NNAM solution looks promising it is used to get initial conditions from equation~(\ref{eq:iv}) for a numerical integration of the equation of motion forward in time, with an order of magnitude smaller time step. In the solutions discussed in Section~\ref{sec:4} present positions and velocities from NNAM and the forward integration never differ by more than 0.2~kpc and 1~km~s$^{-1}$, but for security the latter with the smaller time steps is used. The masses, present positions, and MW circular velocity $v_c$  are iteratively shifted, one at a time, to get a new solution, first from NNAM and then the forward integration. If a parameter shift decreases a $\chi^2$ measure of fit the shift is increased by 25\%\ for the next application. If it increases $\chi^2$ the shift is halved, the sign reversed, the parameter reset to half the distance between original and trial values, and $\chi^2$ recomputed. 

The simple minimization procedure is adequate for the present purpose but I expect ought to be improved for analyses of more complete mass models. It might also be noted that the apparently simpler procedure of adjusting initial conditions for the forward integration instead of final positions for NNAM  proves to be less efficient because $\chi^2$ is a smoother function of present positions.

\begin{table}[htpb]
\centering
\begin{tabular}{lrrrrrr}
\multicolumn{7}{c}{Table 1: Constraints}\\
\tableline\tableline
quantity  & catalog\qquad \ & \qquad 1\ \  & \qquad 2\ \  & \qquad 3\  \  &\qquad  4\  \ & \qquad 5\   \\
\tableline
$v_c$ & $230\pm 10$ km s$^{-1}$ & 233 & 230 & 236 & 236 & 233 \\
\tableline
$m({\rm MW})$ & $15\times 2^{\pm 1}\times 10^{11}m_\odot$  
& 12.4 & 15.1 & 17.8 & 6.5 & 12.8 \\
$m({\rm M31})$ & $30\times 2^{\pm 1}\times 10^{11}m_\odot$  
& 17.2 & 24.0 & 12.3 & 11.5 & 20.7 \\
$m({\rm LMC})$ & $0.6\times 6^{\pm 1}\times 10^{11}m_\odot$ 
& 0.92 &  1.91 & 0.20 & 1.48 &  0.93 \\
$m({\rm M33})$ & $1\times 3^{\pm 1}\times 10^{11}m_\odot$  
& 0.71 & 0.62 & 0.44 & 4.6  & 5.2 \\
$m({\rm IC10})$ & $0.5\times 3^{\pm 1}\times 10^{11}m_\odot$ 
& 0.59 &  0.53 & 0.22 & 0.20 & 0.44  \\
$m({\rm N3109})$ & $1\times 3^{\pm 1}\times 10^{11}m_\odot$  
& 2.3 & 0.54  & 5.4 & 0.58 & 0.48 \\
$m({\rm N300})$ & $2\times 3^{\pm 1}\times 10^{11}m_\odot$ 
 & 2.2 &  4.2 & 2.3 & 1.38 & 0.79  \\
$m({\rm Maffei})$ & $60\times 3^{\pm 1}\times 10^{11}m_\odot$
& 57  & 85  & 87 & $\ldots$ & $\ldots$ \\
\tableline
$cz({\rm M31})$ & $-300\pm 30$ km s$^{-1}$ 
& $-257$ & $-275$  & $-279$ & $-293$ & $-318$  \\
$cz({\rm LMC})$ & $278 \pm 10$ km s$^{-1}$  
 & 278 & 276 & 280 &  280 & 276 \\
$cz({\rm M33})$ & $-179 \pm 20$   km s$^{-1}$ 
& $-196$  & $-192$  & $-219$  & $-227$ & $-206$ \\
$cz({\rm IC10})$ & $-348 \pm 20$   km s$^{-1}$ 
& $-349$ & $-345$  & $-339$ & $-320$ & $-319$ \\
$cz({\rm N3109})$ & $403 \pm 30$   km s$^{-1}$ 
& 358 &  364  & 303 & 392 & 356 \\
$cz({\rm N300})$ & $144 \pm 40$   km s$^{-1}$ 
& 136 & 120  & 134 & 135 & 135 \\
$cz({\rm Maffei})$ & $ 48 \pm 100$  km s$^{-1}$ 
& 25 & 26  & 30 & $\ldots$  & $\ldots$ \\
\tableline
$D({\rm M31})$ & $ 0.785\pm 0.1$ Mpc
& 0.79 & 0.91 & 0.67 & 0.78 & 0.79  \\
$D({\rm LMC})$ & $ 49\pm 6$  kpc
& 53 & 55 & 58 & 50 &  51 \\
$D({\rm M33})$ & $ 0.81\pm 0.1$  Mpc
& 0.74  & 0.68 & 0.88 & 0.75  & 0.67 \\
$D({\rm IC10})$ & $0.76\pm 0.2$  Mpc
& 0.96 & 1.21 & 0.85 & 1.02 & 1.29  \\
$D({\rm N3109})$ & $1.3\pm 0.1$  Mpc
& 1.37 & 1.37 &  1.43 & 1.31 &  1.37 \\
$D({\rm N300})$ & $2.0\pm 0.4$  Mpc
& 2.11 & 2.25 & 2.06 & 2.00 & 2.15 \\
$D({\rm Maffei})$ & $ 3.1\pm 1.0$  Mpc
& 3.1 & 3.4 & 2.7  & $\ldots$ &  $\ldots$ \\
\tableline
$d_\perp({\rm M31})$ &${ \pm 10}$ kpc
& 6.7 &  4.1 & 0.9 & 0.5 &  9.7 \\
$d_\perp({\rm LMC})$ & ${ \pm 2 }$ kpc
& 0.8 &  2.0&   6.4  & 1.3 & 5.1 \\
$d_\perp({\rm M33})$ & ${ \pm 6 }$  kpc
& 1.4 &  1.3 & 1.7 & 1.8 & 2.8 \\
$d_\perp({\rm IC10})$ & ${ \pm  6}$  kpc
& 0.7 &  1.5 & 1.5 & 6.1 & 0.4 \\
$d_\perp({\rm N3109})$ & ${\pm 10 }$  kpc
& 0.3 &  0.3 & 2.0 & 0.6 & 0.4 \\
$d_ \perp({\rm N300})$ & ${ \pm 0.3 }$  Mpc
& 0.06 & 0.12 & 0.22 &  0.10 &0.05 \\
$d_\perp({\rm Maffei})$ & ${ \pm 1}$ Mpc
& 1.3 & 0.5 &  1.2 & $\ldots$ & $\ldots$ \\
\tableline
$\mu_{\alpha}({\rm LMC})$& $ 2.03\pm 0.08$ mas y$^{-1}$
& 2.05 & 2.08  & 2.05 & 1.99 & 2.08 \\
$\mu_{\delta}({\rm LMC})$& $ 0.44\pm 0.05$ mas y$^{-1}$
&  0.44 & 0.44 & 0.47 & 0.42 & 0.43 \\
$\mu_{\alpha}({\rm M33})$& $ 23 \pm 6$ $\mu$as y$^{-1}$
& 22 &  15 &  20 & 22 & 5 \\
$\mu_{\delta}({\rm M33})$& $ 2\pm 7$ $\mu$as y$^{-1}$
& 11 &  8  & 4 & $-4$ & 6 \\
$\mu_{\alpha}({\rm IC10})$& $-2 \pm 8$ $\mu$as y$^{-1}$
& 13& 10  & 9 & 24 & 14 \\
$\mu_{\delta}({\rm IC10})$& $ 20\pm 8$ $\mu$as y$^{-1}$
& 10 & 20 & 16 & $-11$ & $-10$\\
\tableline
$\chi^2$ & $\ldots$ & 25 & 30 & 48 & 46 & 61 \\
\tableline
\end{tabular}
\end{table}

\subsection{Physical parameters}\label{sec:32}

The measure of how well the solutions fit the constraints is the sum
\beq
\chi^2 = \sum ({\rm catalog}_i - {\rm model}_i)^2/\sigma_i^2,
\label{eq:chis}
\eeq
over the parameters that are adjusted --- MW circular velocity, particle masses, and Cartesian components of present particle positions --- and the parameters that are derived --- redshifts, proper motions, and initial peculiar velocities. 

NNAM provides orbits with initial velocities that approximate the growing mode of departure from homogeneity, but that leaves free the magnitude of the initial velocity, which corresponds to the freedom of choice of the amplitude of the growing departure from homogeneity in an ideal fluid. Initial velocities are constrained by assigning $\sigma_v=50$ km s$^{-1}$ to each Cartesian components of the initial peculiar velocity of each galaxy. This choice is motivated by the thought that initial peculiar velocities might be expected to be comparable to velocities within growing protogalaxies. The solutions presented in the next section have reduced $\chi^2\sim 0.3$ for the initial velocities, meaning NNAM produces smaller initial velocities than the adopted estimate. But this term in the total $\chi^2$  is needed  because NNAM also produces solutions with much larger initial velocities. (The sum over $\chi^2$ ignores the fact that the center of mass is at rest, meaning velocities are slightly overcounted, but correcting this does not seem worth the trouble.) 

The cosmological parameters in the Friedmann equation (\ref{eq:FLeq}) are fixed at
\beq
\Omega = 0.27, \qquad H_o = 70\hbox{ km s}^{-1}\hbox{ Mpc}^{-1}.
\eeq
Limited experiments with other choices within the current uncertainties indicate the model results are not sensitive to these parameters.

The second column in Table 1 lists catalog or nominal values of the parameters other than initial velocities in the $\chi^2$ sum. Central values of angular positions, redshifts, and most distances are based on NED. Many of the central values of masses are from an earlier study of the LG and its surroundings (Peebles, Phelps,  Shaya, \& Tully 2001), but the values for LMC, M31 and IC10 are guesses based on luminosities. Most assigned  uncertainties $\sigma_i$ in positions and redshifts are larger than measurement errors.  They may be taken to be estimates of how far the visible galaxies might be offset from the effective centers of the dominant dark matter halos, or as a way to allow for the approximate nature of the model. The quantity $d_\perp$ in the table is the adopted standard deviation in each of the two orthogonal components of the linear distance between catalog and model angular positions (where $d_\perp$ is measured from the catalog position along the line perpendicular to the catalog direction to the line in the angular direction of the model). The masses are imagined to have lognormal distributions, that is, the difference of logarithms of model and catalog masses enters the square in equation~(\ref{eq:chis}). All other parameters are treated as normal variables.   

The first entry in the table is the circular velocity $v_c$ of MW at our position. The central or nominal catalog value is taken to be intermediate between the standard and a possibly larger value (Reid et al. 2009), and the adopted uncertainty allows the standard at one standard deviation. The choice of $v_c$ affects the model for the MW mass distribution (eq.~[\ref{eq:g_LMC}]), but more important is the relation between heliocentric and  galactocentric velocities. The Solar velocity relative to the local standard of rest is taken to be $U=11.1$, $V=12.2$, $W=7.2$ km s$^{-1}$ (Sch{\"o}nrich et al. 2010). This is fixed, not an entry in $\chi^2$. 

The central value for the mass of MW is in the range of usual estimates, with an allowance of twice or half this value at one nominal standard deviation. The effective center of the mass associated with MW is taken to be at the center of the galaxy, fixed at 8.5~kpc distance from the Sun. It would more consistent but overly complicated to allow an offset of this position along with the other galaxies.  The mass of M31 is assigned a larger central value, as is thought to be likely, again with a factor of two allowance at one standard deviation. The uncertainty in the effective redshift of the mass belonging to M31 is taken to be about 10\% of the circular velocity. This seems reasonable but, as for many of the other parameter uncertainties, it is an intuitive guess. The distance is from the survey by McConnachie et al. (2005), but the adopted standard deviation is larger than the stated measurement error. The allowed perpendicular distance $d_\perp$ of the effective center of the mass of M31 from the observed center of the galaxy is 10~kpc, about the optical width of the galaxy. 

The mass of LMC is allowed a larger range because it is interesting to see whether the dynamics give some indication of its likely value. The distance and its uncertainty are from Freedman et al. (2001). van der Marel et al. (2002) describe an uncertainty of about 1~kpc in the position of the optical center of the LMC. The choice $d_\perp=2$~kpc for the offset from the dark matter halo thus seems justified but maybe overly optimistic. It is adopted because an offset  much larger than this significantly shifts the LMC angular position, confusing the meaning of the measured proper motions. The LMC proper motions in Table 1 are from Kallivayalil et al. (2006), and the adopted standard deviations for $\chi^2$ are their stated errors (with no allowance for possible motion of the stars relative to a dark matter halo). 

The galaxies M33 and IC10 are in this analysis because their proper motions (Brunthaler et al. 2005; Brunthaler et al. 2007) are important constraints. The stated uncertainties in the measured proper motions are treated as standard deviations. The allowed perpendicular offset $d_\perp$ is more generous than for the LMC because the angular positions are much less sensitive to $d_\perp$. The larger allowance in the distance to IC10 seems warranted by its low galactic latitude (Sanna et al. 2008; Kim et al. 2009). 

The galaxy NGC3109 is included  because this relatively small spiral with its scattering of dwarf companions is in a low density region on the edge of or just outside LG.  At this location the redshift and distance of NGC3109 is expected to give a particularly direct constraint on the LG mass. The distance is from Dalcanton et al. (2009), with standard deviation larger than the measurement error. 

The object NGC300 is meant to represent the mass belonging to this galaxy and to M55. It may also represent the tidal field of the more distant galaxies in the Sculptor Group. To accommodate this, the allowed ranges of position and redshift are broader than the assignments for the nearer objects. The object Maffei is similarly meant to represent the mass around IC342 and Maffei 1 and perhaps also the tidal effect of more distant mass in roughly the same direction. This first exploration of how the orbit of the LMC relative to the MW might be affected by large mass concentrations external to the LG compares models with and without Maffei. 

\begin{figure}[htpb]
\begin{center}
\includegraphics[angle=0,width=5.25in]{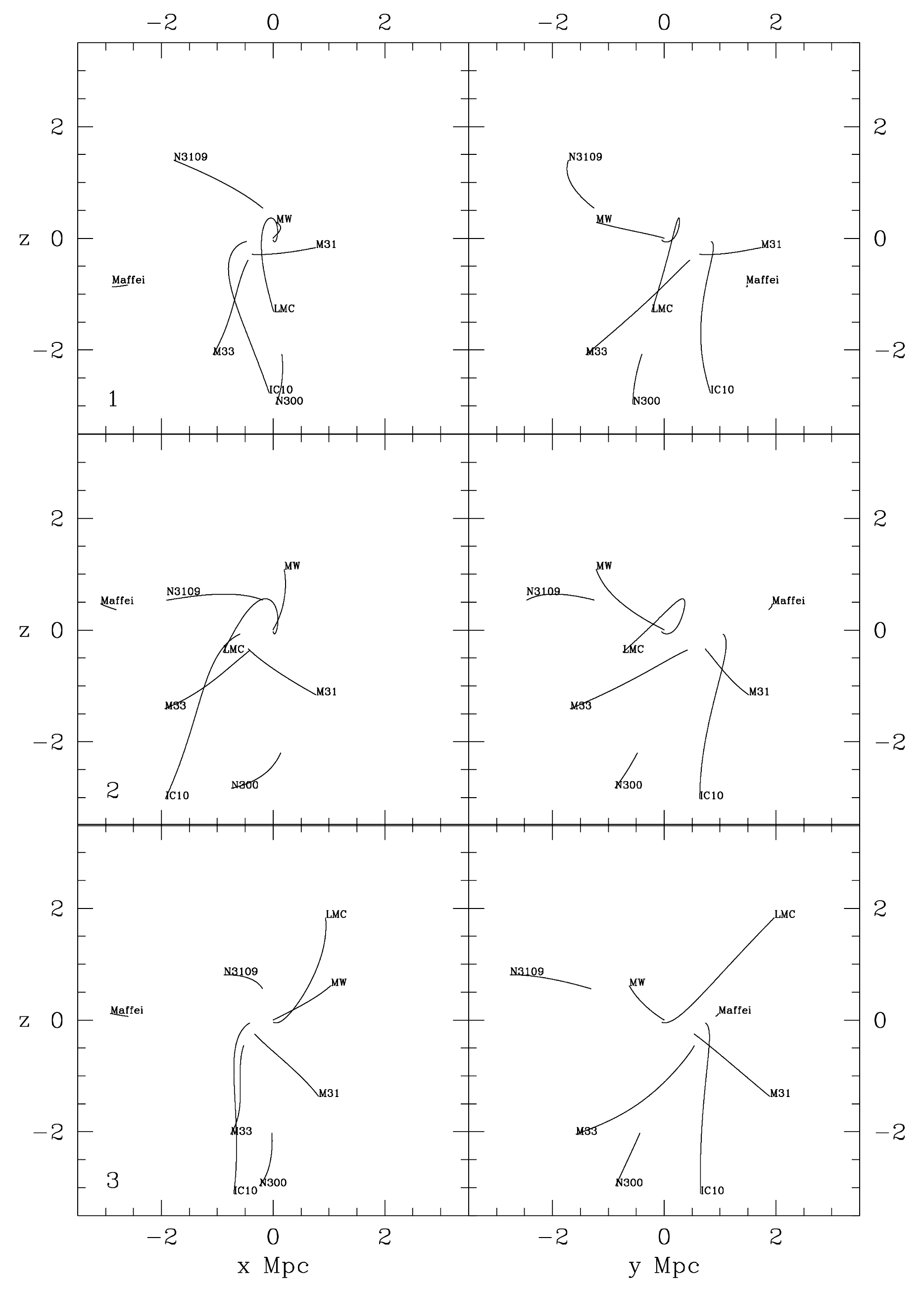} 
\caption{\small Orbits in models 1 to 3 in Table 1 in a right-hand comoving coordinate system with $x$-axis along galactic $l=b=0$ and $z$-axis along $b=90^\circ$. The center of mass is at rest and the origin placed at the present position of the Milky Way. Labels are near initial positions at $1+z_i=10$.\label{Fig:1}}
\end{center}
\end{figure}

\begin{figure}[htpb]
\begin{center}
\includegraphics[angle=0,width=5.25in]{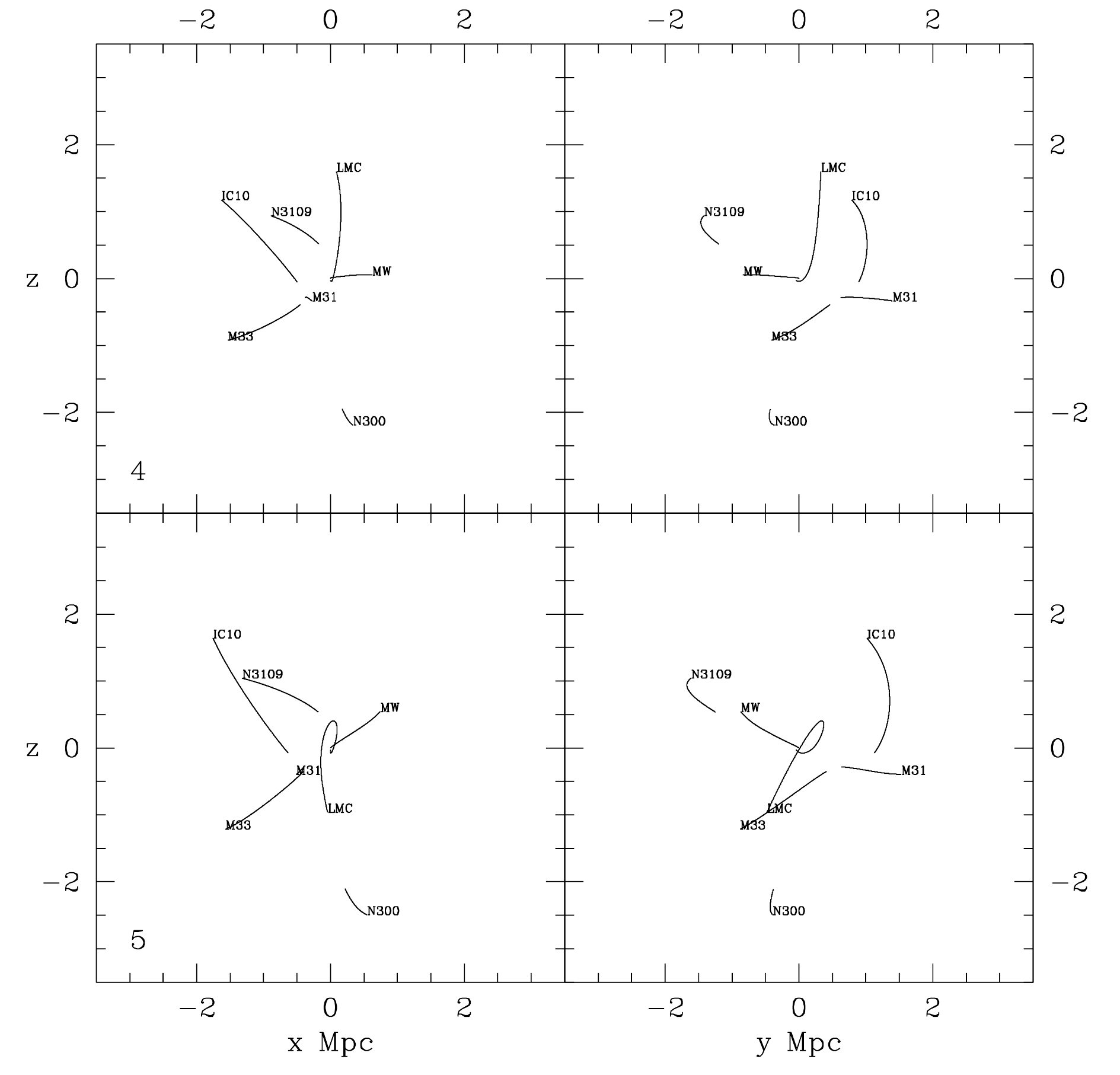} 
\caption{\small Same as Figure \ref{Fig:1} for models 4 and 5 in Table 1. \label{Fig:2}}
\end{center}
\end{figure}

\section{Models}\label{sec:4}

Table 1 lists parameters in three solutions that include the large external mass in the direction of the Maffei system of galaxies, and two solutions without this external mass. All 
five models fit the measured motion of LMC. The $\chi^2$ measure of fit to all the constraints is entered in the bottom row of Table~1. It is the sum over squares of differences of model and nominal values of $v_c$, 8 masses, 7 redshifts, $3\times 2$ proper motions, $7\times 3$ components of present positions, and  $7\times 3$ initial velocities, for a total of 64 parameters that are imagined to have Gaussian distributions. The 30 quantities adjusted to reduce $\chi^2$ are $v_c$, the 8 masses, and the $7\times 3$ components of present positions. Thus we might expect $\chi^2\sim 34$ in a realistic solution, if the uncertainties $\sigma_i$ were realistic.

The particle orbits in comoving galactic coordinates are shown in Figures~\ref{Fig:1} and~\ref{Fig:2}. They show two types of motion of LMC and MW. Type I, in Models 1, 2 and 5, has LMC initially moving up in a clockwise direction in the $xz$ and $yz$ projections. Type II, in Models 3 and 4, has LMC more directly approaching MW from above with a sharp bend at the low redshift end. These two general orbit types, with considerable variations in the orbits of M33 and IC10, are identifiable in all solutions I have found with $\chi^2\la 60$. The two orbit types largely differ by the motion of MW relative to the center of mass of the mass model --- the motion of LMC relative to MW is quite similar, as will be discussed --- but there are two possibly significant points to note. First, parameter adjustments in the walk to lower $\chi^2$ for Type II solutions often end at an abrupt change of orbits and a large increase in $\chi^2$. The walk in Type I seems to be approaching a smooth minimum that is not quite reached in acceptable computation time. Second, limited trials with the initial expansion factor changed from equation~(\ref{eq:zi}) to $a_i=0.2$ yielded Type II solutions that are very close to what is found at $a_i=0.1$,  but did not yield Type I solutions. 

Models 1 and 2 have the smallest $\chi^2$ found in this study. They have similar  parameters, and the orbits are similar but distinctly different. Both are shown to illustrate the variations allowed in similar arrangements of the orbits at different apparently local minima of $\chi^2$. Models 3 and 4 are included to show the alternative Type II behavior. Models 4 and 5 illustrate the effect of eliminating the large external mass, at about the lowest $\chi^2$ for this purpose.

All 5 models put $v_c$ above 220 km~s$^{-1}$. This may not be significant, however, because the nominal catalog value is larger too, and may have biased the choice of local minima. Remaining to be done is a search for solutions based on a smaller catalog value of $v_c$. 

Since the masses are allowed considerable departures from nominal it is a positive result that in Models 1 and 2 the mass of M31 is larger than MW, as is usually considered likely. The sums of masses of M31 and MW are $3\times 10^{12}m_\odot$ and  $4\times 10^{12}m_\odot$, a substantial difference largely due to the allowed difference of model redshifts of M31. In these two models the components of proper motion of IC10 differ from the measurement by less than two times the stated error, which seems acceptable, and the other proper motions are closer to the measurements. The distance of IC10 in model 2 is large, but obscuration complicates this measurement.  These models place Maffei well away from its catalog position, at $d_\perp=1.3$ and $0.5$~Mpc, but this is an acceptable outcome of the idea that Maffei may  represent external mass in the general direction of the nearest large galaxy concentration. In both models $\chi^2$ is consistent with what is expected from the parameter count. But since many of the nominal catalog standard deviations are not much better than guesses the conclusion is that the models are reasonable fits to what is thought to be known within the ambiguities proposed in Section~\ref{sec:32}. 

Model 3 has a reasonably small $\chi^2$ but three questionable features. First, it puts LMC near the edge of the optical image, at $d_\perp =6.4$~kpc. Second, MW is more massive than M31, which seems unlikely though perhaps not impossible. Third, the redshift of NGC3109 is 100 km~s$^{-1}$ below the catalog value, which again seems unlikely. To be checked is whether a more complete mass model would allow the Type II pattern of motion to better fit arguably reasonable  constraints. 

Models 4 and 5, without the large external mass, have acceptable proper motions of LMC and proper motions of M33 that are arguably not unreasonable. The proper motion of IC10 is off by nearly four standard deviations, but this is may be acceptable for the purpose of modeling the motion of LMC,  because at its greater distance IC10 may have been affected by more distant objects not in this mass model. Perhaps the greatest objection to Models 4 and 5 is that they replace the external mass with a large mass of M33, at roughly half the mass of the MW, which seems unlikely. But the important issue for the present study is how this rearrangement of the more distant mass affects the orbit of LMC relative to MW.

\begin{figure}[htpb]
\begin{center}
\includegraphics[angle=0,width=3.1in]{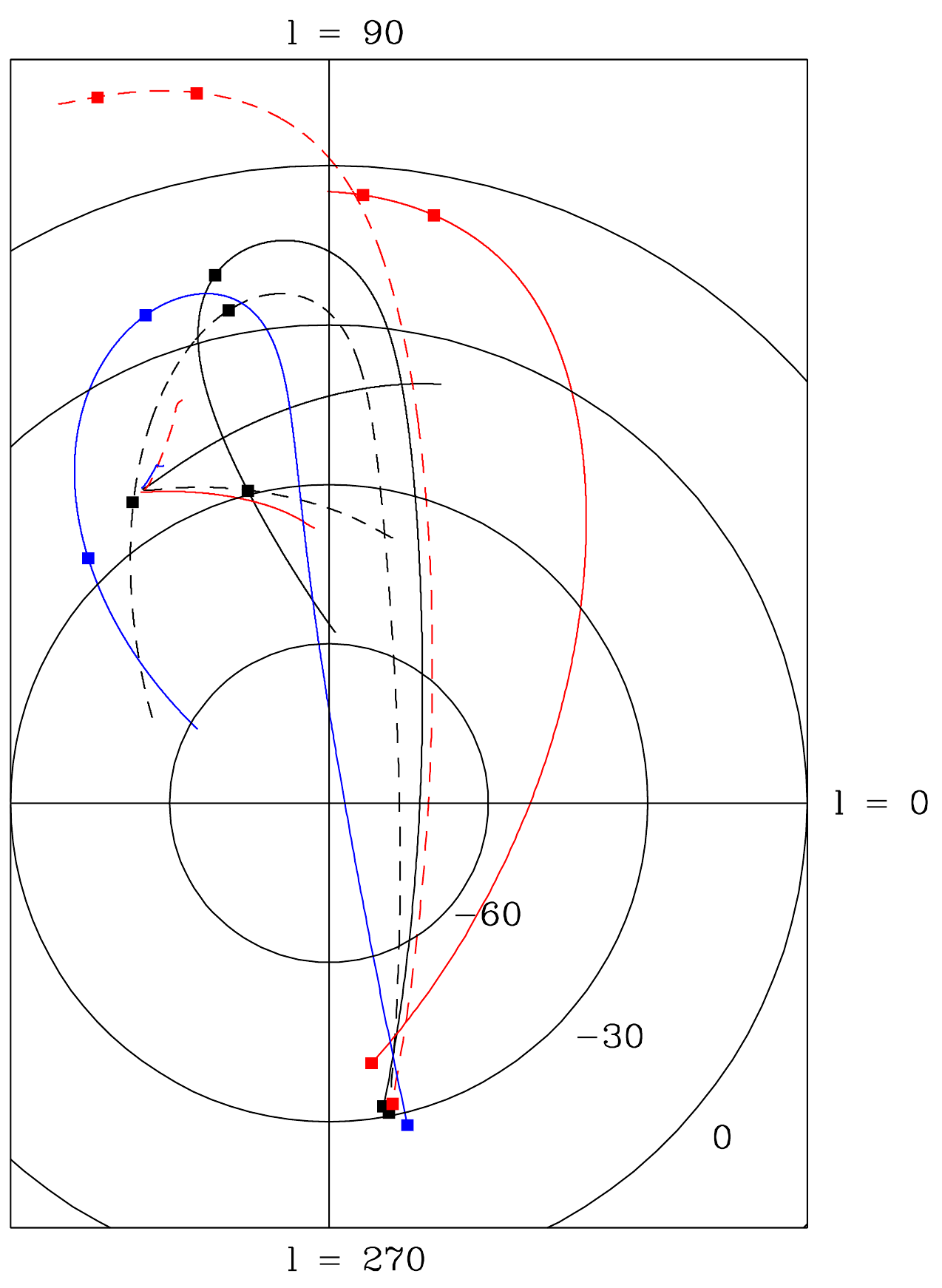} 
\caption{\small Angular positions of LMC and M31 relative to the present solar position, for  Model 1 (solid black), 2 (dashed black), 3 (solid red), 4 (dashed red) and 5 (blue).  The curves converge on present positions of LMC near the bottom of the figure and of M31 near the left-hand side where the shorter curves converge. The squares on LMC orbits mark positions at $z = 0$, 1, and 3. \label{Fig:3}}
\end{center}
\end{figure}

\begin{figure}[htpb]
\begin{center}
\includegraphics[angle=0,width=5.5in]{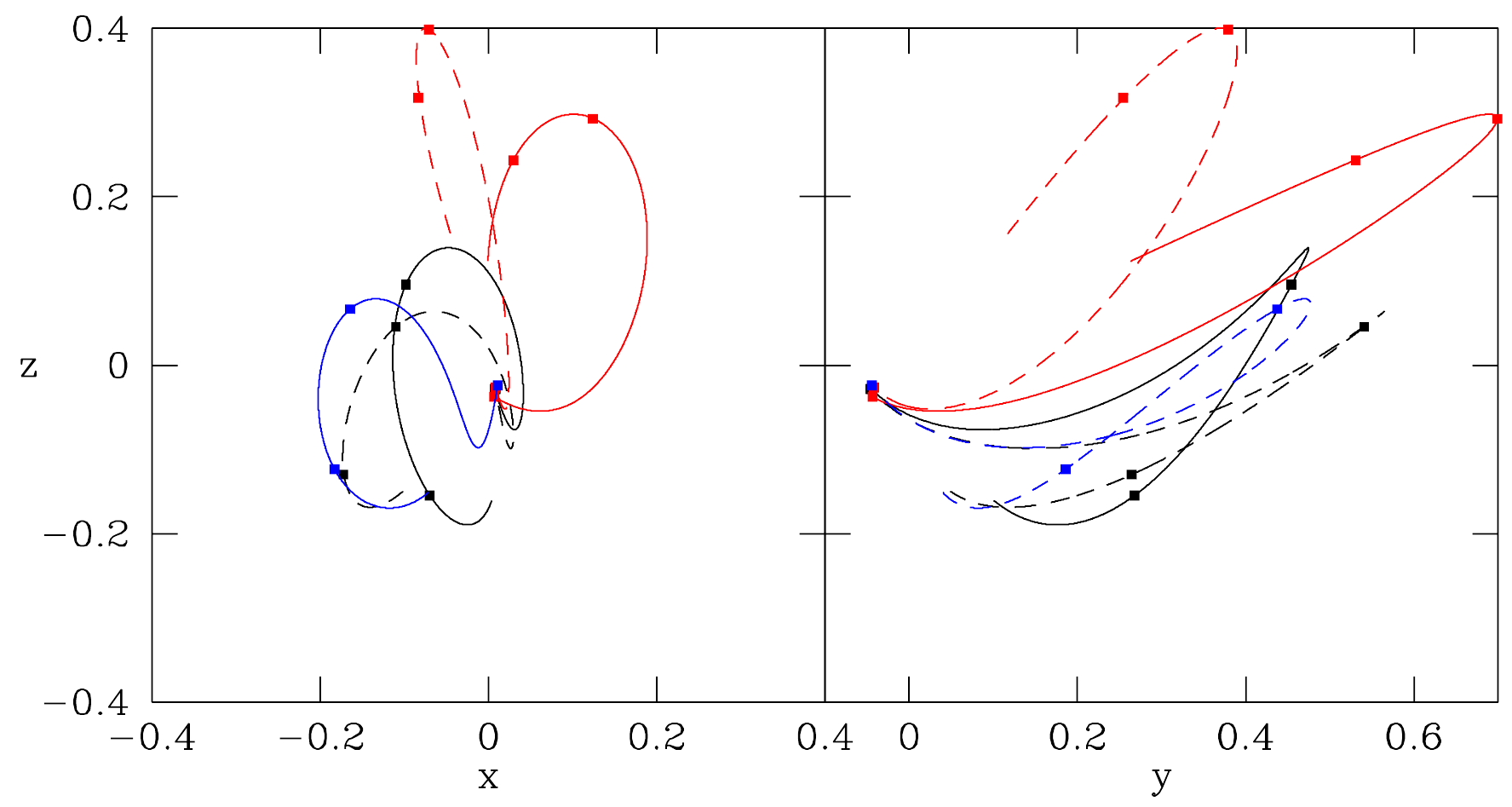} 
\caption{\small Orbits of LMC relative to the present solar position in the coordinate system in Figures~\ref{Fig:1} and~\ref{Fig:2} with the labeling scheme in Figure \ref{Fig:3}. \label{Fig:4}}
\end{center}
\end{figure}

Figure \ref{Fig:3} shows the history of angular positions of LMC and M31 for an observer fixed in the MW at the present position of the Sun in the galaxy. The present angular position of LMC is near the bottom of the figure. Model 3, with its questionable value of $d_\perp$, is most different from the other orbits. The considerable scatter among the others includes a clear difference between the orbits of the two most plausible Models 1 and 2 (plotted in black). The orbits share some distinctive features, however. All except Model 3 passed within $\sim 20^\circ$ of the South galactic pole, in the general direction of the Magellanic stream (Mathewson \& Ford 1984). Within the scatter these models agree with the Besla et al. (2007, Fig. 8) model for the motion of LMC past the pole. At redshift near unity all five orbits pass a stationary point in latitude near $b=0$. At this point the longitude is near $l=90^\circ$ and decreasing. This behavior is seen also in the more schematic mass model with cosmological initial conditions in P9. 

The present position of M31 is toward the left side of Figure \ref{Fig:3}. In all models M31 is moving to increasing galactic longitude, in the opposite direction to the motion of LMC. The net displacement of M31 is smallest in Models 4 and 5 (and the blue curve is so short it is difficult to see in the figure), as might be expected from the absence of torquing by a  large external mass. The heliocentric velocity of M31 is $v_\alpha=234$ km~s$^{-1}$, $v_\delta=-65$ km~s$^{-1}$ in Model 1 and $v_\alpha=223$ km~s$^{-1}$, $v_\delta=8$ km~s$^{-1}$ in Model 2.  van der Marel \& Guhathakurta (2008) argue for $v_\alpha=78\pm 41$ km~s$^{-1}$, $v_\delta=-38\pm 34$ km~s$^{-1}$, which differs by nearly four times the stated error. Understanding this discrepancy requires further discussion of both approaches, and, it is to be hoped, the situation will be established by direct proper motion measurements in M31. 

Figure \ref{Fig:4} gives a three-dimensional picture of the orbits of LMC relative to MW at the present Solar position. One sees again the considerable differences but also the similarity of the patterns of motion. In Model 1, LMC is 190 kpc away from MW and 300~kpc from M31 at $1+z=10$, and LMC reaches maximum distance 500~kpc from MW at $z=0.66$. The numbers are quite similar in Model 2. Given the MW model mass, the initial and maximum LMC-MW separations are similar to that of a radial orbit that leaves an isolated MW at high redshift and has just returned. That is, the orbit is largely dominated by MW, though the other galaxies are needed to account for the development of this orbit out of cosmological initial conditions.

\section{Conclusions}

The case that  Models 1 and 2 usefully approximate the past motion of the Large Magellanic Cloud relative to the Milky Way (in the black curves in Figs.~\ref{Fig:3} and~\ref{Fig:4}) commences with the existence of these arguably plausible fits to the parameters in Table 1. This result was not guaranteed, and it is important therefore that the search has yielded one --- and only one --- fully acceptable arrangement of orbits, with the variations illustrated by the differences between Models 1 and 2. The values of $\chi^2$ in these two models are consistent with the counts of constraints and adjustable parameters within the nominal standard deviations in Table 1. Since most of these standard deviations are based on intuitive arguments this result should be read to mean that Models 1 and 2  are plausible within the arguments of reasonableness in Section~\ref{sec:3}. One could attempt to develop a more quantitative assessment of the meaning of the values of $\chi^2$ by applying the analysis to random catalogs, but that does not seem worthwhile; the case is better made by broader considerations. 

One such consideration is that the nature of the orbit of  LMC relative to MW is similar in all five models (Figs.~\ref{Fig:3} and~\ref{Fig:4}). These models all fit the measured motion of LMC, and they generally fit all the measured redshifts and positions in the mass model  within the assigned uncertainties. Model 3 has problems with the larger mass of MW, the large offset of LMC from the optical center, and the small redshift of NGC3109, but it shares the single close passage of LMC and the  swing to larger galactic longitude going backward in time at redshift about unity. Models 4 and 5 illustrate the need for three massive actors in addition to LMC to account for the motion of LMC out of the LMC-MW-M31 plane.  (As noted in P9, the gravitational interaction among three galaxies, LMC, MW and M31, in an otherwise empty universe drives motions confined to the plane of the three galaxies.) The models without a large external mass solve the problem by promoting M33 to a massive actor. This is unlikely but a useful illustration of what the measured LMC motion requires. And it is to be noted that the LMC orbits relative to MW in Models 4 and 5 share the general features of the other solutions. The same is true of the still more schematic model in P9. 

Since the motion of LMC at low redshifts is dominated by the mass in MW, it is not surprising that the orbits in Figure~\ref{Fig:3} at low redshift all are similar and agree with models with MW alone (Besla et al. 2007) or with MW and M31 (Kallivayalil et al. 2009; Shattow  \& Loeb 2009). The swing toward larger galactic longitude at  larger redshift, and the single close approach of LMC to MW, are less intuitive, but they have proved to be stable results of the measured LMC motion under the cosmological initial condition that the primeval peculiar velocities are growing. 

 Besla et al. (2007) and Shattow \& Loeb (2009) present models in which LMC has completed more than one orbit around MW. This is not found in the present analysis. The random starting orbits were designed to reach plausible solutions with two close passages,  but the lack of discovery of examples  must be balanced against the experience that the action method is not well adapted to capturing this case. The argument from astronomy against an earlier close passage is that it might be expected to have stripped away the H{\small I} (van den Bergh 2006). Thus Besla et al. (2010) find that the single orbit allows a model for the origin of the Magellanic stream by tidal interaction between the Large and Small Magellanic Clouds. 
  
The computation allows considerable freedom in the mass of LMC, and indeed the mass in Model 3 is one tenth that of Model 2. However, the more plausible Models~1 and~2 put the mass of LMC at 1 to $2\times 10^{11}m_\odot$. This is ten times the mass within 10 kpc (van der Marel et al. 2002), but at luminosity $L_B\sim 3\times 10^9L_\odot$ a conventional dark matter halo could make up the difference. Indeed, Besla et al. (2010) present an independent argument for an LMC mass $\sim 2\times 10^{11}m_\odot$ from the relation between luminosity and halo mass found in numerical simulations of structure formation in the standard cosmology. If LMC had a massive halo as it approached MW then this dark matter would now be far from smoothly distributed as the halo merges with the MW halo. A halo merger does not seem likely to have had a significant effect on the motion of the stellar part of LMC as it plunges into the dominant MW dark halo, but a numerical analysis to check this would be interesting.  

These arguments make a good --- though not definitive --- case that the orbit of LMC relative to MW has the shape illustrated in Figures~\ref{Fig:3} and~\ref{Fig:4}. An issue to be revisited with better mass models is that this analysis depends on the assumption that the Magellanic Clouds originated as clumps of matter left little disturbed by the assembly of the major galaxies. An alternative picture in which the Clouds are debris from a violent major merger at a more modest redshift (Yang \& Hammer 2010) is not likely to be found by NNAM. This picture will have to be evaluated by other considerations, including observational and theoretical studies of the baryonic debris from mergers.

The difference between the LMC orbits in the two plausible models (plotted in black) shows that under the cosmological initial condition there still is considerable uncertainty in where LMC was at redshift $z=1$. This might be reduced by tighter modeling of what more distant galaxies were doing. 

Though models 1 and 2 are reasonable fits to the measured proper motions of M33 and IC10 as well as LMC, the M33 orbit does not agree with the proposal by McConnachie et al. (2009) that M33 passed close to M31. Since the orbits of these two galaxies are sensitive to the orbit of M31 a firmer case for where M33 and IC10 have been awaits a firmer case for the proper motion of M31. In all five models in Table 1, M31 is moving toward increasing longitude, and in the two most plausible cases the motion to increasing right ascension is faster than the van der Marel \& Guhathakurta (2008) estimate. Since the transverse motion of M31 is expected to be more sensitive than LMC to the mass distribution outside LG a firmer case for the proper motion of M31 and the orbits of M33 and IC10 awaits more detailed mass models that more completely account for the large galaxies exterior to the Local Group and, equally important, include  more of the nearby isolated smaller galaxies that largely serve as test particles. The analysis presented here shows that NNAM will be be well suited for analyses of these more ambitious mass models, though likely with a more efficient way to minimize the $\chi^2$ measure of fit and a better strategy to search for acceptable orbits, perhaps along the lines of Peebles et al. (2001). Most important, of course, will be the tighter constraints from advances in measurements of galaxy distances and proper motions, ground-based and from the Gaya and SIM lite satellite missions. 

This analysis is based on a picture for the mass distribution and a cosmology that have passed searching tests. But the cosmology and our ideas about how mass is distributed around galaxies still are enormous extrapolations from what actually is well established. The apparently successful fit to the motions of the very nearby galaxies is a modest but not trivial addition to our fund of cosmological tests, and a test that can and should be considerably improved.

\acknowledgments

I have benefitted from advice from Ed Shaya and Brent Tully. This research has made use of the NASA/IPAC Extragalactic Database (NED) which is operated by the Jet Propulsion Laboratory, California Institute of Technology, under contract with the National Aeronautics and Space Administration. The numerical matrix inversion of equation~(A15) is from Press et al. (1992).

\appendix
\section{New Numerical Action Method}

This describes the NNAM solution of equation (\ref{eq:dreamon}) for the adjustment of the coordinates of a single particle toward a solution of the equation of motion. Finding  simultaneous shifts of all coordinates of all particles follows by the same argument. It is not presented here because I have found shorter computation times for iterated adjustments of coordinates of one particle at a time.

Let
\beq
F^+_n=  
{a_{n+1/2}^2\dot a_{n+1/2}\over a_{n+1} - a_n}, 
\quad
 F^-_n=  
{a_{n-1/2}^2\dot a_{n-1/2}\over a_{n} - a_{n-1}}
= F^+_{n-1}, \quad {dt_n\over a_n} = {t_{n+1/2}-t_{n-1/2}\over a_n},
\label{eq:F}
\eeq
and for simplicity drop the label for the particle $i$ whose position is being adjusted. Then equation (\ref{eq:discrete_eom}) is
\beq
S_{k,n} = -F^+_n(x_{k,n+1}-x_{k,n}) +  F^-_n(x_{k,n}-x_{k,n-1})
+ {dt_n\over a_n}\left( g_{k,n}  + 
{1\over 2}\Omega H_o^2 x_{k,n}\right),\label{eq:eom}
\eeq
with $F^-_1=0$. 

The equation to be solved for the coordinate shifts $\delta x_{k,n}$ of the particle whose orbit is being adjusted is 
\beq
S_{k,n} + \sum_{k',n'} S_{k,n,k',n'}\delta x_{k',n'} = 0, \qquad 
{\partial^2 S\over\partial x_{k,n}\partial x_{k',n'}} = S_{k,n;k',n'} = S_{k',n';k,n}.
\label{eq:tosolve}
\eeq
The nonzero derivatives are
\beq
S_{k,n;k,n+1} = - F^+_n, \qquad S_{k,n;k,n-1} = - F^-_n, 
\eeq
\beq
S_{k,n,k',n} =
(F^+_n + F^-_n)\delta_{k,k'} + {dt_n\over a_n}\left(\sum_{j\not= i}{\partial g_{k,n}\over\partial x_{k',n}} + {\Omega H_o^2\over 2}\delta_{k,k'}\right). 
\eeq
Since $S_{k,n;k',n'} = 0$ unless $n'=n$ or else $n' = n\pm 1$ and $k' = k$, equation (\ref{eq:tosolve}) is
\beq
S_{k,n} + S_{k,n;k, n+1}\delta x_{k,n+1}+ 
 \sum_{k'}S_{k,n;k',n}\delta x_{k',n}
+ S_{k,n;k,n-1}\delta x_{k,n-1}= 0.  \label{eq:tosolvea}
\eeq
Set $n\rightarrow n - 1$ in this equation and rearrange it to get
\beq
\delta x_{k,n} = - {S_{k,n-1}+ \sum_{k'} S_{k,n-1;k',n-1}\delta x_{k',n-1} 
+ S_{k,n-1;k,n-2}\delta x_{k,n-2}
\over S_{k,n-1;k,n} }. \label{eq:tosolveb}
\eeq
This gives $\delta x_{k,n}$ in terms of $\delta x_{k,n-1}$ and $\delta x_{k,n-2}$. On iterating we arrive at the form
\beq
\delta x_{k,n} = A_{k,n} + \sum_{k\pp} B_{k,n;k\pp}\delta x_{k\pp,1}.\label{eq:iteq}
\eeq
At $n=1$ this is just
\beq
 A_{k,1} = 0,\qquad B_{k,1;k\pp} = \delta_{k,k\pp}.
\eeq
At the second time step from the start, $n=2$, equation~ (\ref{eq:tosolveb}) is
\beq
\delta x_{k,2} = - \bigg[S_{k,1}+ \sum_{k'} S_{k,1;k',1}\delta x_{k',1} \bigg]/
 S_{k,1;k,2} , \label{eq:forneq2}
\eeq
because there is no $\delta x_{k,0}$. Comparing this with equation~ (\ref{eq:iteq}) we see that
\beq
A_{k,2} = - S_{k,1}/S_{k,1;k,2},
\qquad B_{k,2;k'} = - S_{k,1;k',1}/S_{k,1;k,2}.
\label{coefficients}
\eeq
At $n\geq 3$ the result of substituting the form (\ref{eq:iteq}) into equation~ (\ref{eq:tosolveb})  is
\beqa
\delta x_{k,n} = &-& \bigg[S_{k,n-1} + \sum_{k'} S_{k,n-1;k',n-1}(A_{k',n-1} 
+ \sum_{k\pp} B_{k',n-1;k\pp}\delta x_{k\pp,1}) \nonumber\\
&+& S_{k,n-1;k,n-2}(A_{k,n-2} + \sum_{k\pp} B_{k,n-2;k\pp}\delta x_{k\pp,1})\bigg]
/ S_{k,n-1;k,n}. \label{eq:34}
\eeqa
This shows that at $3\leq n\leq n_x$ the constants in equation~(\ref{eq:iteq}) are
\beqa
&&A_{k,n} = - {S_{k,n-1} + \sum_{k'} S_{k,n-1;k',n-1}A_{k',n-1}
+ S_{k,n-1;k,n-2}A_{k,n-2}\over S_{k,n-1;k,n}},\nonumber \\
&& B_{k,n;k\pp} = - {\sum_{k'} S_{k,n-1;k',n-1}B_{k',n-1;k\pp}
+ S_{k,n-1;k,n-2}B_{k,n-2;k\pp}\over S_{k,n-1;k,n}}. \label{fixcoeffs}
\eeqa

Equations (\ref{coefficients}) and (\ref{fixcoeffs}) give the $A_{k,n}$ and $B_{k,n;k\pp}$, $2\leq n\leq n_x$, in terms of the given derivatives of the action. Then equation~~(\ref{eq:iteq}) is $3n_x - 3$ equations for the $\delta x_{k,n}$, at $2\leq n\leq n_x$, in terms of the $\delta x_{k,1}$. We get three more, which fix the $\delta x_{k,1}$, by setting $n=n_x$ in equation (\ref{eq:tosolvea}) and recalling that $\delta x_{k,n_x+1}= 0$:
\beqa
0 &=& S_{k,n_x}  +  \sum_{k'}S_{k,n_x;k',n_x}\delta x_{k',n_x}
+ S_{k,n_x;k,n_x-1}\delta x_{k,n_x-1} \nonumber\\
&=&  S_{k,n_x}  +  \sum_{k'}S_{k,n_x;k',n_x}
\bigg[ A_{k',n_x} + \sum_{k\pp} B_{k',n_x;k\pp}\delta x_{k\pp,1}\bigg] \\
&& \qquad\ + S_{k,n_x;k,n_x-1}\bigg[
A_{k,n_x-1} + \sum_{k\pp} B_{k,n_x-1;k\pp}\delta x_{k\pp,1}\bigg], \nonumber
\eeqa
or
\beq
0 = T_k + \sum_{k'}T_{k,k\pp}\delta x_{k\pp,1}, \label{eq:15}
\eeq
where
\beqa
&& T_k = S_{k,n_x}  +  \sum_{k'}S_{k,n_x;k',n_x}A_{k',n_x} 
+ S_{k,n_x;k,n_x-1}A_{k,n_x-1}, \nonumber\\
&& T_{k,k\pp} = \sum_{k'}S_{k,n_x;k',n_x}B_{k',n_x;k\pp}
+ S_{k,n_x;k,n_x-1} B_{k,n_x-1;k\pp}. \label{eq:16}
\eeqa
This is three equations for the three $\delta x_{k,1}$. The rest of the $\delta x_{k,n}$ then follow from equations (\ref{eq:iteq}) and~(\ref{fixcoeffs}). 

For $N_p$ particles and $n_x$ time steps the computation time to get the $\delta x_{i,k,n}$  for all particles scales as $N_p^2N_x$. In previous applications of the numerical action method (P9 and references therein) equation~(\ref{eq:tosolve}) is solved by matrix inversion, an operation that scales as $N_p^2n_x^3$ for relaxation of one particle orbit at a time. The NNAM operation scales in the same way as a conventional numerical integration, but the numerical prefactor is considerably larger because it takes many iterations of the coordinate shifts $\delta x_{i,k,n}$ to drive the $S_{i,k,n}$ to zero. NNAM certainly will not replace conventional forward numerical integration for simulations of the growth of cosmic structure. But the application presented here shows why NNAM will be useful for analyses of the dynamics of the nearby galaxies.

\end{document}